\documentclass[prl,twocolumn]{revtex4}
\usepackage{amsmath}
\usepackage{amsfonts}
\usepackage{amssymb}
\usepackage{graphicx}%

\begin{document}

\title{Flexible Quasi-Three-Dimensional Terahertz Metamaterials}

\author{Abul K. Azad,$^{1,*}$ Hou-Tong Chen,$^{1}$ Antoinette J. Taylor,$^{1}$ Elshan Akhadov,$^{1}$ Nina R. Weisse-Bernstein,$^2$ and John F. O'Hara$^{1}$}
\affiliation{$^1$MPA-CINT, Los Alamos National Laboratory, P. O. Box 1663, MS K771, Los Alamos, New Mexico 87545, USA}
\affiliation{$^2$ISR-2: Space and Remote Sensing, Los Alamos National Laboratory, P. O. Box 1663, MS B244, Los Alamos, NM 87545, USA}
\affiliation{$^*$Corresponding author: aazad@lanl.gov}

\begin{abstract} We characterize planar electric terahertz metamaterials
fabricated on thin, flexible substrates using terahertz time-domain
spectroscopy.  Quasi-three dimensional metamaterials are formed by stacking
multiple metamaterial layers.  Transmission measurements
reveal resonant band-stop behavior that becomes stronger with an
increasing number of layers.  Extracted metamaterial dielectric
functions are shown to be independent of the number of layers,
validating the effective medium approximation.  Limitations of
this approximation are discussed.
\end{abstract}

\maketitle

Electromagnetic metamaterials enable numerous exotic responses not
available in natural materials, such as negative refractive
index~\cite{Shelby_Science_2001}, perfect
lensing~\cite{Penrdy_PRL_2001}, and cloaking
~\cite{Schurig_Science_2006}. They typically consist of
sub-wavelength metallic resonators fabricated on dielectric or
semiconducting substrates and can be made to respond to either or
both of the electric and magnetic
fields of incident waves. By proper scaling, individual
resonators and the composite metamaterial can resonantly respond
to electromagnetic waves of nearly any frequency, upto optical.
Typically, the size of the resonators is roughly an order of
magnitude smaller than the operating wavelength.
Split-ring-resonators (SRRs) in combination with wire arrays,
initially introduced by Pendry, have been widely used for
metamaterial designs. These structures present a considerable
fabrication challenge at higher frequencies ($\geq100$~GHz) and
therefore microwave techniques have dominated experimental
metamaterial research. This is especially relevant for
bulk, or three-dimensional (3D), metamaterials, where
multi-layered structures become necessary.

High frequency, multi-layer metamaterials can be fabricated by dicing and
stacking single resonator arrays grown on thin substrates
 ~\cite{Gokkavas_PRB_2006, BDF_APL_2007}.  Or they can be monolithically
grown on a single substrate using multi-layer processing
~\cite{Katsarakis_OL_2005,NA_NatureMaterial_2007}. The latter
process is particularly suitable for optical metamaterials where
the spacing between layers is only in the sub-micron or micron
range.  Either approach works for terahertz (0.1-10 THz)
metamaterials though the different spacing between successive
layers creates new challenges.  An additional challenge, at all
frequencies, is to create flexible or conformable metamaterials,
which would be particularly useful in certain applications, such
as cloaking or shielding.  Fabrication techniques using rigid or
fragile substrates, such as FR-4 or silicon, are clearly unfit.
However, some progress has been made here.  Multi-layer
metamaterials have been demonstrated in the far-infrared frequency
regime by using polyimide fillers between resonator layers
~\cite{Katsarakis_OL_2005}.

In the past few years, metamaterials research has sparked special
interest in the THz community.  Metamaterials are an optimistic
approach to overcome the limitations of natural materials in the
construction of novel functional THz devices, some of which have
recently been demonstrated ~\cite{HTC_Nature_2006,
Padilla_PRL_2006, HTC_OL_2007}. Planar SRR arrays fabricated on
various semiconductor and insulator substrates have proven capable
of responding to THz radiation electrically ~\cite{Azad_OL_2006}
and/or magnetically ~\cite{XZhang_Science_2004, DRSmith_APL_2007}.
Other designs provide an approach to negative permittivity
~\cite{HTC_Nature_2006} or permeability
~\cite{XZhang_Science_2004}.

In this work, we address the issues of flexibility, 3D
fabrication, and material parameter extraction by studying the
transmission properties of THz metamaterials comprised of metallic
electric split-ring-resonators (eSRRs) on conformable polyimide
(Kapton) substrates using terahertz time domain spectroscopy
(THz-TDS). These are studied as single layers and also as quasi-3D
THz metamaterials by stacking up to four layers. In all cases, the
measured transmission exhibits a clear minimum at 1.12 THz because
of the excitation of the inductor-capacitor (LC) resonance. The
transmission evolves into a full stop-band centered at the
resonance frequency as the number of layers increases in the
stacked quasi-bulk medium. Complex dielectric properties of the
single and multiple layer samples are extracted and we show how
they relate to the effective medium approximation.

\begin{figure}
\begin{center}
\includegraphics[ width=2.5in,keepaspectratio=true]{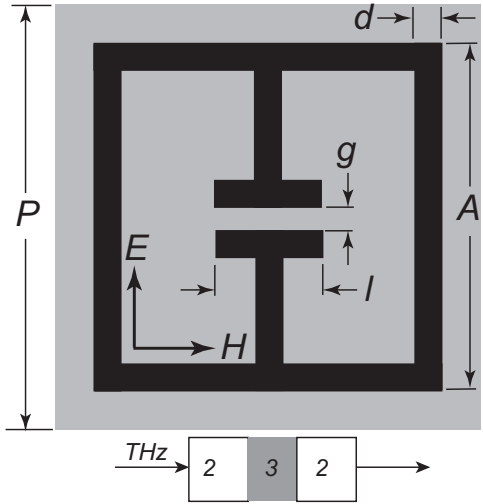}
\caption{Diagram of split-ring-resonator unit cell.
Black area represents the metallized resonator and gray area
represents the unit cell. E and H shows the orientations of
incident THz electric and magnetic fields. Bottom shows the
schematic of the THz propagation through the metamaterial (region
3) sandwiched between the quartz plates (region 2).}

\end{center}
\end{figure}

The eSRR structure employed in this work is shown in Fig.~ 1. The
square resonators of length $A=40\ \mu$m are arranged in a square
lattice of period $P=54\ \mu$m with line width $d=3\ \mu$m,
capacitor plate length $l=12\ \mu$m, and gap $g=3\ \mu$m. Using
standard photolithography and e-beam evaporation, the eSRRs are
fabricated in a square array on a commercially available polyimide
film with a measured thickness of 84 $\pm$ 2~$\mu$m.  Polyimide
has been widely used for photonic and electronic devices because
of its high electrical and thermal stability~\cite{CPWong_Kapton}.
Its flexibility, durability, relatively low refractive index
($n\cong2$) and absorption ($\alpha\cong 20$~cm$^{-1}$) also make
it favorable as a THz metamaterial substrate.  Metallization
consists of 10~nm of titanium followed by 200~nm of gold. During
photolithography a regular silicon wafer is attached to the
polyimide film to provide the mechanical support. The
polyimide-based single layer metamaterials were then cut, visually
aligned, and tightly stacked to form the quasi-3D media. All
samples had a 10~mm~$\times$~10~mm patterned active area.

THz transmission measurements were performed with wave incidence
normal to the eSRR plane in a confocal, photoconductive antenna
based THz-TDS system~\cite{JFO_APL_2006}. The system generates
broadband pulsed THz radiation with a frequency independent beam
waist of 3~mm diameter at the sample. Samples are placed in THz
path with the gap bearing arm of the eSRR oriented parallel to the
THz electric field, as shown in Fig.~1. The time-domain THz fields
are recorded following passage through the samples and a dry air
reference. All of the samples are sandwiched between two 1~mm
thick quartz plates during measurements.  Quartz has an index
roughly equivalent to polyimide, which ensures that the front
metamaterial layer has similar boundary conditions as subsequent
layers.  The reference is measured with the quartz plates in
contact with each other. The complex sample $E_{S}(\omega)$ and
reference $E_{R}(\omega)$ spectra are calculated by numerical
Fourier transform. The effective material parameters of the sample
can then be extracted from the measured transmission function
$t(\omega)=E_{S}(\omega)/E_{R}(\omega)$ via ~\cite{Azad_APL_2003},

\begin{equation}
 t(\omega)=t_{23}t_{32}\frac{\exp\big(ik_0d(\tilde{n}-1)\big)}{1+
 r_{23}r_{32}\exp(i2k_0d\tilde{n})}\nonumber
\end{equation}

\noindent where $t_{23}$, $t_{32}$ and $r_{23}$, $r_{32}$ are the
frequency-dependent complex transmission and reflection
coefficients through the interfaces at the quartz-metamaterial boundaries;
$k_0$ is the
free-space wavenumber, $d$ is the sample thickness, and
$\tilde{n}=n+i\kappa$ is the sample complex refractive index.  Time-windowing
allows us to disregard multiple reflections in the quartz plates.

\begin{figure}
\begin{center}
\includegraphics[width=3.4 in,keepaspectratio=true]
{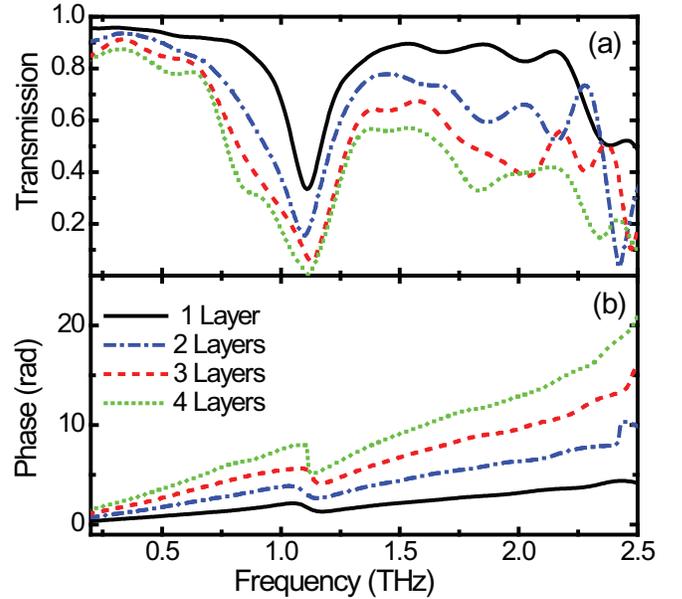} \caption{(a) Measured transmission amplitude of THz
electric field through metamaterial samples with different number
of layers. (b) The corresponding phase change. The measurements of
1-layer, 2-layer, 3-layer, and 4-layer metamaterials are
represented by solid (black), dashed-dot (blue), dashed (red), and
dotted (green) curves, respectively.}

\end{center}
\end{figure}

Figure 2(a) shows the measured THz amplitude following
transmission through the metamaterials.  Measured transmission
spectra reveal clear resonances at 1.12~THz due to the effective
inductive-capacitive (LC) response of the eSRRs.  Transmissions
were also measured through the multiple layer samples and then
compared with the single layer case.  The transmission spectra
show band-stop behavior that increases in strength with an
increasing number of layers. At four layers the effective medium
becomes completely band-stop at the resonance frequency whereas
the single layer transmission minimum is 33\%. The low frequency
off-resonance transmission also decreases from 95\% in a single
layer metamaterial to 85\% in a four-layer metamaterial. The
overall linewidth of the LC resonance increases with number of
layers, though there is no significant shift in the resonance
center frequency. This demonstrates that the resonance frequency
is mainly determined by the SRR structural parameters.  We also
note the ripple features, most noticeable between the LC
(1.12~THz) and dipolar (2.5~THz) resonances. Simulations initially
indicate that these are due to the multiple reflections within or between
sample layers. This is under investigation, but we
presently restrict our discussion to the behavior of the LC
resonance, our main interest.

Figure 2(b) shows the phase changes incurred by the THz pulses
during transmission through the metamaterials. Both the substrate
polyimide and the metallic eSRRs contribute to this phase change:
the linear contribution comes from the substrate while the
resonance structure comes from eSRRs. As expected, the slope of
the linear contribution increases as the wave encounters more
substrate.

\begin{figure}
\begin{center}
\includegraphics[width=3.4 in,keepaspectratio=true]
{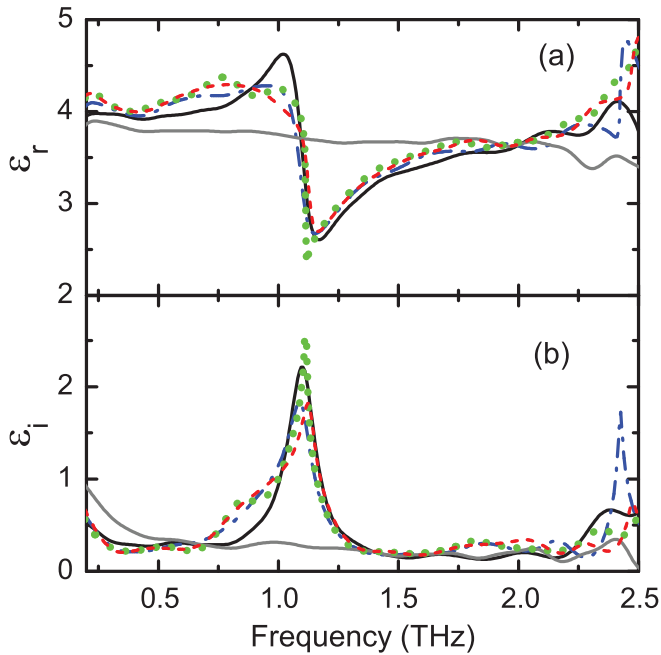} \caption{Extracted (a) real and (b) imaginary
dielectric functions of the effective metamaterials. They are
presented using the same patterns (colors) as in Fig. 2. The solid
gray curve represents the dielectric function of the bare
polyimide film. }

\end{center}
\end{figure}

The optical properties of the samples can be extracted from the
measurements, but are critically dependent on the choice of some
effective metamaterial thickness. Often this effective thickness
is defined by enforcing a cubic unit cell for each resonator
~\cite{Schurig_APL_2006}.  In our case this would be 54~$\mu$m.
Since our substrate is 84~$\mu$m thick, this approach introduces
complexity in the analysis and can even cause incorrect results
[16].  Instead, we simply assign each metamaterial layer the same
thickness as our substrate, 84~$\mu$m.  The complex dielectric
function is computed from the measured complex index by
$\epsilon(\omega)=\epsilon_{r}(\omega)+i
\epsilon_{i}(\omega)=\tilde{n}^2$. Figures.~3(a) and (b) show
$\epsilon_r$ and $\epsilon_i$, respectively, for the individual
and stacked metamaterials, and the bare substrate.  The bare
polyimide properties are fairly constant over our frequency range.
The metamaterial properties are largely independent of the number
of layers, in all the cases showing a fairly flat $\epsilon_r$
averaging about 3.8 with an approximately Lorentzian perturbation
of about $\pm1$ around resonance.  The off-resonance metamaterial
dielectric functions are very similar to those of the substrate
with deviations that could be due to loss in the metal structures
and tolerances in the substrate thickness.

Figures 2 and 3 reveal how our metamaterials behave in light of
the effective media approximation.  Simply expressed,
$\epsilon(\omega)$ defines a material's electric response (e.g.
phase shift) as a function of wave propagation distance.  Thus,
for homogenous media having fixed $\epsilon(\omega)$, thicker
samples obviously produce larger cumulative responses.  Our
multi-layer samples create larger responses than the single layer
(see Fig.~2), but these are distributed over greater thicknesses
(more layers), yielding an unchanging $\epsilon(\omega)$.  This
behavior is consistent with an effective medium approximation.
Though not detailed here, this approximation fails at the
individual layer.  For any single layer the entire resonant
response is produced by the very thin SRR array
\cite{JFO_APEC_2007}; it does not accumulate in passage through
the substrate.  For waves in transmission this fixed response is
effectively distributed over the entire substrate thickness.
Therefore, thicker substrates dilute the response, and ultimately
flatten the Lorentzian variation in $\epsilon(\omega)$.  This
behavior is obviously not characteristic of a truly homogeneous
medium and marks an important limitation of the effective medium
treatment when applied to planar metamaterials.

In conclusion, we have demonstrated THz metamaterials based on
eSRRs arrays fabricated on flexible, thin polyimide substrates and
compared the transmission properties of the single-layer and
quasi-3D bulk metamaterials using THz-TDS. Our results further
verify a fabrication approach by which truly bulk 3D, durable, and
conformable metamaterials can be realized; these are obviously
important steps in realizing functional THz metamaterial devices
such as prisms, lenses, waveguides, filters, and cloaks or
shields.  Finally, our results illustrate interesting details
about the proper application of the effective medium approximation
as it applies to planar metamaterials.  This work was performed,
in part, at the Center for Integrated Nanotechnologies, a U. S.
Department of Energy, Office of Basic Energy Sciences nanoscale
science research center jointly operated by Los Alamos and Sandia
National Laboratories. Los Alamos National Laboratory, an
affirmative action equal opportunity employer, is operated by Los
Alamos National Security, LLC, for the National Nuclear Security
Administration of the U. S. Department of Energy under contract
DE-AC52-06NA25396. We gratefully acknowledge the support of the U.S.
Department of Energy through the LANL/LDRD Program for this work.

\end{document}